\documentclass[twocolumn,twoside,slac]{revtex4}
\usepackage{url}
\usepackage{hyperref}
\usepackage{graphicx}
\usepackage{fancyhdr}
\newcommand{\globus}{Globus Toolkit$\mathrm{^{TM}}$}

\begin{document}

\title{The NorduGrid architecture and tools}

\author{P.\ Eerola, B.\ K\'{o}nya, O.\ Smirnova}
\affiliation{Particle Physics, Institute of Physics, Lund University, Box 118, 22100 Lund, Sweden}
\author{T.\ Ekel\"of, M.\ Ellert}
\affiliation{Department of Radiation Sciences, Uppsala University, Box 535, 75121 Uppsala, Sweden}
\author{J.\ R.\ Hansen, J.\ L.\ Nielsen, A.\ W\"a\"an\"anen}
\affiliation{Niels Bohr Institutet for Astronomi, Fysik og Geofysik, Blegdamsvej 17, Dk-2100 Copenhagen \O, Denmark}
\author{A.\ Konstantinov} 
\affiliation{University of Oslo, Department of Physics, P.\ O.\ Box 1048, Blindern, 0316 Oslo, Norway and Vilnius University, Institute of Material Science and Applied Research, Saul\.etekio al.\ 9, Vilnius 2040, Lithuania}
\author{F.\ Ould-Saada}
\affiliation{University of Oslo, Department of Physics, P.\ O.\ Box 1048, Blindern, 0316 Oslo, Norway}

\begin{abstract}
  The NorduGrid project designed a Grid architecture with the primary goal to
  meet the requirements of production tasks of the LHC experiments. While it is
  meant to be a rather generic Grid system, it puts emphasis on batch
  processing suitable for problems encountered in High Energy Physics. The
  NorduGrid architecture implementation uses the \globus{} as the foundation
  for various components, developed by the project. While introducing new
  services, the NorduGrid does not modify the Globus tools, such that the two
  can eventually co-exist. The NorduGrid topology is decentralized, avoiding
  a single point of failure.  The NorduGrid architecture is thus a
  light-weight, non-invasive and dynamic one, while robust and scalable,
  capable of meeting most challenging tasks of High Energy Physics.
\end{abstract}

\maketitle

\thispagestyle{fancy}

\section{Introduction\label{sec:intro}}

The European High Energy Physics community is in the final stages of
construction and deployment of the Large Hadron Collider (LHC) -
the world biggest accelerator, being built at the European Particle Physics
Laboratory (CERN) in Geneva. Challenges to be faced by physicists are
unprecedented. Four experiments will be constructed at the LHC to observe
events produced in proton-proton and heavy ion collisions. Data collected by
these experiments will allow for exploration of new frontiers of the
fundamental laws of nature, like the Higgs mechanism with possible discovery
of the Higgs boson; CP-violation in B-meson decays; supersymmetry; large
extra dimensions and others. One of the greatest challenges of the LHC
project will be the acquisition and analysis of the data. When, after a few
years of operation, the accelerator will run at its design luminosity, each
detector will observe bunch collisions at a rate of $4\cdot 10^7$ per second. A
set of filter algorithms, implemented in hardware and on state-of-art
programmable processors, aims to reduce the event rate to less than 1000
events per second for final storage and analysis. The equivalent data volume
is between 100 MByte/sec and 1 GByte/sec. Each experiment is expected to
collect 1 PByte of raw data per year. The two LHC general purpose
experiments, ATLAS and CMS, have each more than 150 participating institutes
distributed all over the world. $2000$ physicists per experiment contribute
to the development of hardware and software and they expect to have almost
instantaneous access to the data and to a set of up-to-date analysis tools.

\subsection{The NorduGrid Project}

In order to face the computing challenge of the LHC and other similar
problems emerging from the science communities the NorduGrid project also
known as the {\it Nordic Testbed for Wide Area Computing} was setup. The goal
was to create a GRID-like infrastructure in the Nordic countries (Denmark,
Norway, Sweden and Finland) that could be used to evaluate GRID tools by the
scientists and find out if this new technology could help in solving the
emerging massive data and computing intensive tasks emerging. The first test
case was to be the ATLAS Data Challenge (see~\cite{nordugrid-atlas}) which
was to start out in May of 2002.

As the focus was on deployment we hoped that we could adopt existing
solutions on the marked rather than develop any software in the project. The
available Grid middleware providing candidates were narrowed down to the
Globus Toolkit\texttrademark{} and the software developed by the European
DataGrid project (EDG)~\cite{edg}\cite{globus}. The middleware from these two
projects however was soon found to be inadequate. The Globus Toolkit is
mainly a box of tools rather than a complete solution. There is no job
brokering and their Grid Resource Allocation Manager~\cite{gram} lacked the
possibility of staging large input and output data. The EDG software seemed
to address the deficiencies of the Globus Toolkit, but were not mature enough
in the beginning of 2002 to seriously solve the problems faced by the ATLAS
data challenges. Furthermore it was felt that the implementation of their
centralized resource broker was a serious bottleneck that could never be used
in solving the large amounts of data that were going to be dealt with on a
large data Grid.

It was therefore decided that the NorduGrid project would create a Grid
architecture from scratch. The implementation would use existing pieces of
working Grid middleware and develop the missing pieces within the project in
order to create a production Grid testbed.

\section{Architecture\label{sec:architecture}}

The NorduGrid architecture was carefully planned and designed to satisfy the
needs of users and system administrators simultaneously. These needs can be
outlined as a general philosophy:

\begin{itemize}
\item Start with simple things that work and proceed from there
\item Avoid architectural single points of failure
\item Should be scalable
\item Resource owners retain full control of their resources
\item As few site requirements as possible:
  \begin{itemize}
  \item No dictation of cluster configuration or install method
  \item No dependence on a particular operating system or version
  \end{itemize}
\item Reuse existing system installations as much as possible
\item The NorduGrid middleware is only required on a front-end machine
\item Compute nodes are not required to be on the public network
\item Clusters need not be dedicated to Grid jobs
\end{itemize}

\subsection{The NorduGrid System Components\label{sec:components}}

\begin{figure*}[!t]
\centering{
\includegraphics[width=0.9\linewidth]{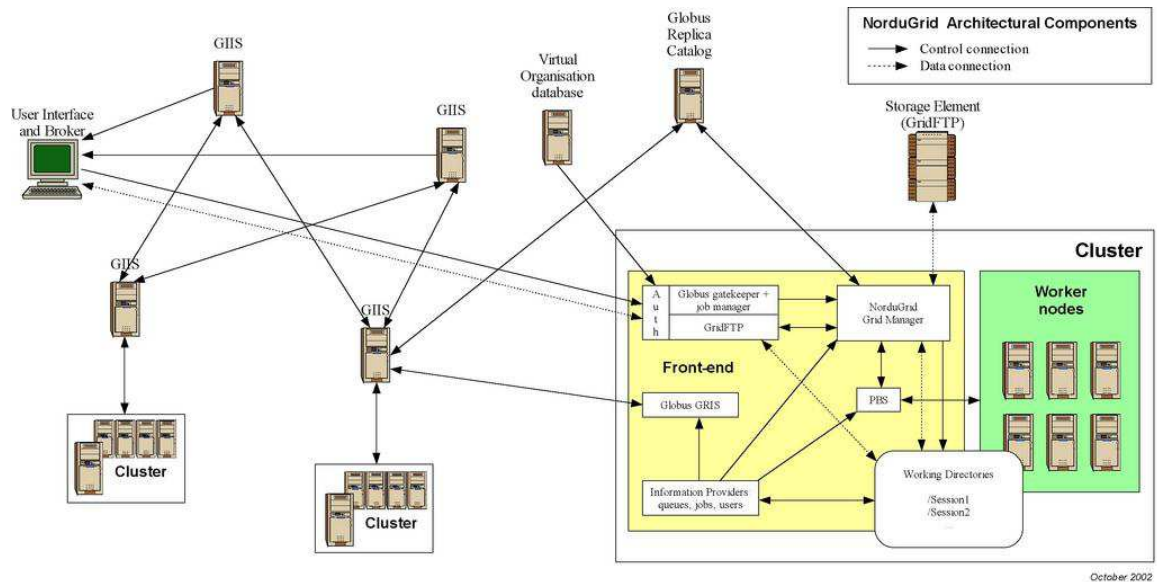}
\caption{\label{fig:nordarch}Components of the NorduGrid architecture.}}
\end{figure*}

The NorduGrid tools are designed to handle job submission and
management, user area management, some data management, and monitoring.
Figure~\ref{fig:nordarch} depicts basic components of the NorduGrid
architecture and schematic relations between them. In what follows,
detailed descriptions of the components are given.

\textbf{User Interface} (UI) is the major new service developed by the
NorduGrid, and is one of its key components. It has the high-level
functionality missing from the \globus, namely, that of resource
discovery, brokering, Grid job submission and job status querying. To
achieve this, the UI communicates with the NorduGrid Grid Manager (see
Section~\ref{sec:gm}) and queries the Information System
(Section~\ref{sec:infosys}) and the Replica Catalog
(Section~\ref{sec:replica}).

The UI comes simply as a client package (Section~\ref{sec:ui}), which
can be installed on any machine -- typically, an end-user desktop
computer. Therefore, NorduGrid does not need a centralized resource
broker, instead, there are as many User Interfaces as users find
convenient.

\textbf{Information System} within the NorduGrid is another key
component, based on MDS~\cite{mds} and realized as a distributed service
(Section~\ref{sec:infosys}), which serves information for other
components, such as a UI or a monitor (Section~\ref{sec:monitor}).

The Information System consists of a dynamic set of distributed
databases which are coupled to computing and storage resources to
provide information on the status of a specific resource. When they
are queried, the requested information is generated locally on the
resource (optionally cached afterward), therefore the information
system follows a pull model. The local databases register themselves
via a soft-state registration mechanism to a set of indexing services
(registries). The registries are contacted by the UI or monitoring
agents in order to find contact information of the local databases for
a direct query.

\textbf{Computing Cluster} is the natural computing unit in the
NorduGrid architecture. It consists of a front-end node which manages
several back-end nodes, normally via a private closed network. The
software component consists of a standard batch system plus an extra
layer to interface with the Grid. This layer includes the Grid Manager
(Section~\ref{sec:gm}) and the local information service
(Section~\ref{sec:infosys}).

The operation system of choice is Linux in its many flavors. Other
Unix-like systems (\textit{e.g.}  HP-UX, Tru64 UNIX) can be utilized
as well.

The NorduGrid does not dictate configuration of the batch system,
trying to be an "add-on" component which hooks the local resources
onto the Grid and allows for Grid jobs to run along with the
conventional jobs, respecting local setup and configuration policies.

There are no specific requirements for the cluster setup, except that there
should be a shared file system (\textit{e.g.} The Network File System - NFS)
between the front-end and the back-end nodes. The back-end nodes are managed
entirely through the local batch system, and no NorduGrid middleware has to
be installed on them.

\textbf{Storage Element} (SE) is a concept not fully developed by the
NorduGrid at this stage. So far, SEs are implemented as plain
GridFTP~\cite{gridftp} servers. A software used for this is a GridFTP
server, either provided as a part of the \globus, or the one delivered
as a part of the NorduGrid Grid Manager (see
Section~\ref{sec:gm}). The latter is preferred, since it allows access
control based on the user's Grid certificate instead of the local
identities to which users are mapped.

\textbf{Replica Catalog} (RC) is used for registering and locating
data sources. NorduGrid makes use of the RC as developed by the Globus
project~\cite{globus}, with minor changes to improve functionality
(see Section~\ref{sec:replica}). RC records are entered and used
primarily by the GM and its components, and can be used by the UI for
resource brokering.

\subsection{Task Flow\label{sec:task}}

The components described above are designed to support the task flow
as follows:
\begin{enumerate}
\item A user prepares a job description using the extended
  Globus Resource Specification Language (RSL,
  Section~\ref{sec:xrsl}). This description may include application
  specific requirements, such as input and output data description, as
  well as other options used in resource matching, such as
  architecture or an explicit \textbf{cluster}.
\item The job description is interpreted by the \textbf{UI}, which
  makes resource brokering using the \textbf{Information System} and
  \textbf{RC} data, and forward the job to the chosen cluster (see
  Section~\ref{sec:ui}), eventually uploading specified accompanying
  files.
\item The job request (in extended RSL format) is received by the
  \textbf{GM} which resides on cluster's front-end. It handles
  pre-processing, job submission to the local system, and
  post-processing (see Section~\ref{sec:gm}), depending on the job
  specifications. Input and output data manipulations are made by the
  GM with the help of \textbf{RC}'s and \textbf{SE}'s.
\item A user may request e-mail notifications about the job status, or
  simply use the UI or monitor (Section~\ref{sec:monitor}) to follow
  the job progress. Upon the job end, specified in the job description
  files can be retrieved by the user. If not fetched in 24 hours, they
  are erased by the local GM.
\end{enumerate}

In~\cite{nordugrid-atlas} this can be seen in a real production environment.

\section{The NorduGrid Middleware}

The NorduGrid middleware is almost entirely based on the \globus \ API,
libraries and services. In order to support the NorduGrid
architecture, several innovative approaches were used, such as the
Grid Manager (Section~\ref{sec:gm}) and the User Interface
(Section~\ref{sec:ui}). Other components were extended and developed
further, such as the Information Model (Section~\ref{sec:infosys})
and the Extended Resource Specification Language
(Section~\ref{sec:xrsl}).

\subsection{Grid Manager} \label{sec:gm}

The NorduGrid Grid Manager (GM) software acts as a smart front-end
for job submission to a cluster. It runs on the cluster's
front-end and provides job management and data pre- and
post-staging functionality, with a support for meta-data
catalogs.

The GM is a layer above the \globus \ libraries and services. It was
written because the services available as parts of the \globus \ did
not meet the requirements of the NorduGrid architecture
(Section~\ref{sec:architecture}) at that time. Missing things included
integrated support for the RC, sharing cached files among users,
staging in/out data files, \emph{etc.}.

Since data operations are performed by an additional layer, the data are
handled only at the beginning and end of a job. Hence the user is
expected to provide complete information about the input and output
data. This is the most significant limitation of the used approach.

Differently from the Globus Resource Allocation Manager (GRAM)~\cite{gram},
the GM uses a GridFTP interface for jobs submission. To do that, a
NorduGrid GridFTP server (GFS) was developed, based on the Globus
libraries~\cite{gridftp}. Its main features which
makes it different from the Globus implementation are:
\begin{itemize}
\item Virtual directory tree, configured per user.
\item Access control, based on the Distinguished Name stored in the
user certificate.
\item Local file system access, implemented through loadable plug-ins
(shared libraries). There are two plug-ins provided with GFS:
\begin{itemize}
\item The local file access plug-in implements an ordinary FTP
server-like access,
\item The job submission plug-in provides an interface for submission
and control of jobs handled by the GM.
\end{itemize}
\end{itemize}
The GFS is also used by NorduGrid to create relatively easily
configurable GridFTP based storage servers (often called Storage
Elements).

The GM accepts job-submission scripts described in Globus RSL
(Section~\ref{sec:xrsl}) with a few new attributes added.

For every job, the GM creates a separate directory (the \emph{session
directory}) and stores the input files into it. There is no single point
(machine) that all the jobs have to pass, since the gathering of input
data is done directly on a cluster front-end.

Input files local to the user interface are uploaded by the UI. For remote
input files the GM start a download process that understand a variety of
protocols: (eg. http, ftp, gridftp, rc, rsl, etc.).

Once all input files are ready the GM creates a job execution script and
launches it using a Local Resource Management System (LRMS). Such a script
can perform additional actions, such as setting the environment for third-party
software packages requested by a job.

After a job has finished, all the specified output files are transferred
to their destinations, or are temporarily stored in the session
directory to allow users to retrieve them later.

\begin{figure*}[ht]
\centering
\includegraphics[width=0.95\linewidth]{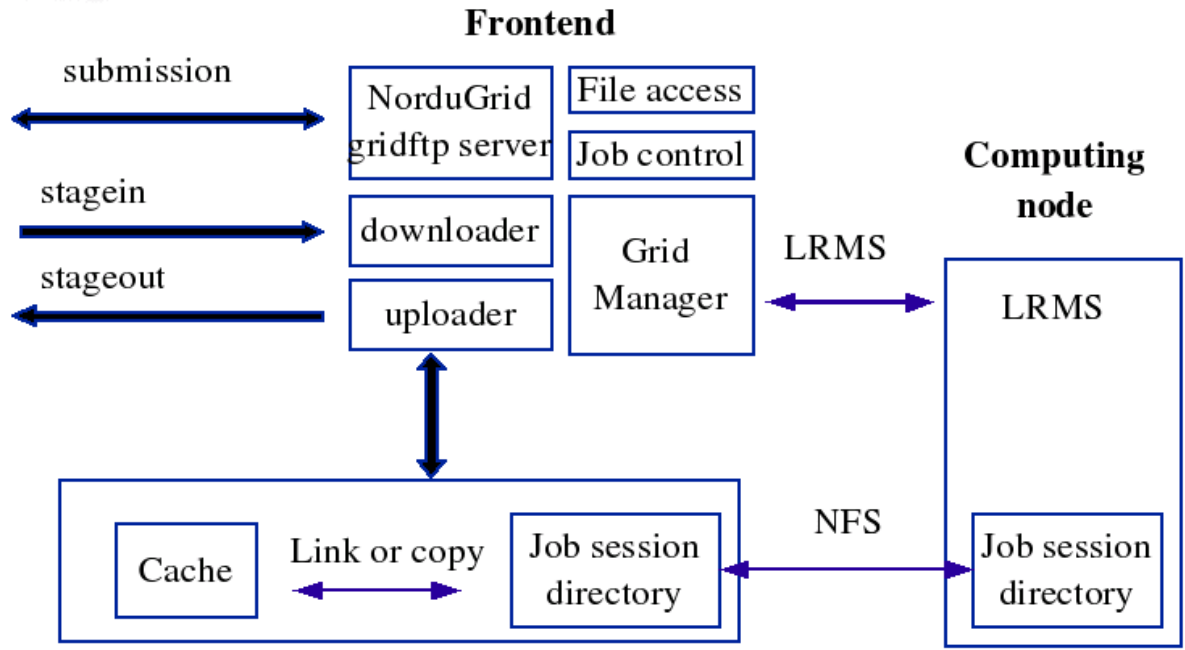}%
\caption{\label{fig:gridmanager}Grid Manager architecture}
\end{figure*}

If an output destination specifies a RC additional registration is performed
in the replica catalog. A schematic view of the GM mechanism is shown on
Figure~\ref{fig:gridmanager}.

Additionally, the GM can cache input files and let jobs and users
share them. If the requested file is available in the cache, it is
still checked (if the protocol allows that) against the remote server
whether a user is eligible to access it. To save the
disk space, cached files \emph{can} be provided to jobs as soft links.

The GM uses a \texttt{bash}-script Globus-like implementation of an
interface to a local batch system. This makes it easy to adapt the GM
to inter-operate with most LRMSs.

The GM uses the file system to store the state of handled jobs. This
allows it to recover safely from most system faults, after a restart.

The GM also includes user utilities to handle data transfer and
registration in meta-data catalogs.

\subsection{Replica Catalog} \label{sec:replica}

The NorduGrid makes use of the Globus RC, which is based on an
OpenLDAP~\cite{ldap} server with the default LDAP DatabaseManager
back-end. There are no significant changes introduced into the
original Globus RC objects schema. Apparent OpenLDAP problems with
transferring relatively big amounts of data over an
authenticated/encrypted connection were fixed partially by applying
appropriate patches, and partially by automatic restart of the
server. Together with a fault tolerant behavior of the client part,
this made the system usable.

To manage the information in the RC server, the \globus \ API and
libraries were used. The only significant change was to add the
possibility to perform securely authenticated connections based on the
Globus GSI mechanism.

\subsection{Information System} \label{sec:infosys}

The NorduGrid Information System description exceeds the limits
imposed by this publication.

The NorduGrid middleware implements a dynamic distributed information
system~\cite{is} which was created by extending the Monitoring and
Discovery Services~\cite{mds} (MDS) of the \globus. The MDS is an
extensible framework for creating Grid information systems, and it is
built upon the OpenLDAP~\cite{ldap} software. An MDS-based information
system consists of an information model (schema), local information
providers, local databases, soft registration mechanism and
information indices. NorduGrid extensions and a specific setup, made
MDS a reliable backbone of the information system. A detailed account
of these modifications is given below.

\textbf{A Grid information model} should be a result of a
delicate design process of how to represent the resources and what is
the best way to structure this information. In the used MDS-based
system, the information is being stored as attribute-value pairs of
LDAP entries, which are organized into a hierarchical tree. Therefore,
the information model is technically formulated via an LDAP schema and
the specification of the LDAP-tree.

Evaluation of the original MDS and the EDG
schemas~\cite{mds,edg-schema} showed that they are not suitable for
describing clusters, Grid jobs and Grid users simultaneously. Therefore,
NorduGrid designed its own information model~\cite{is}. While Globus
and other Grid projects keep developing new schemas~\cite{glue,cim},
the NorduGrid one is the only which has been deployed, tested and used
for a production facility.

\begin{figure*}
{\centering
\includegraphics[width=0.95\linewidth]{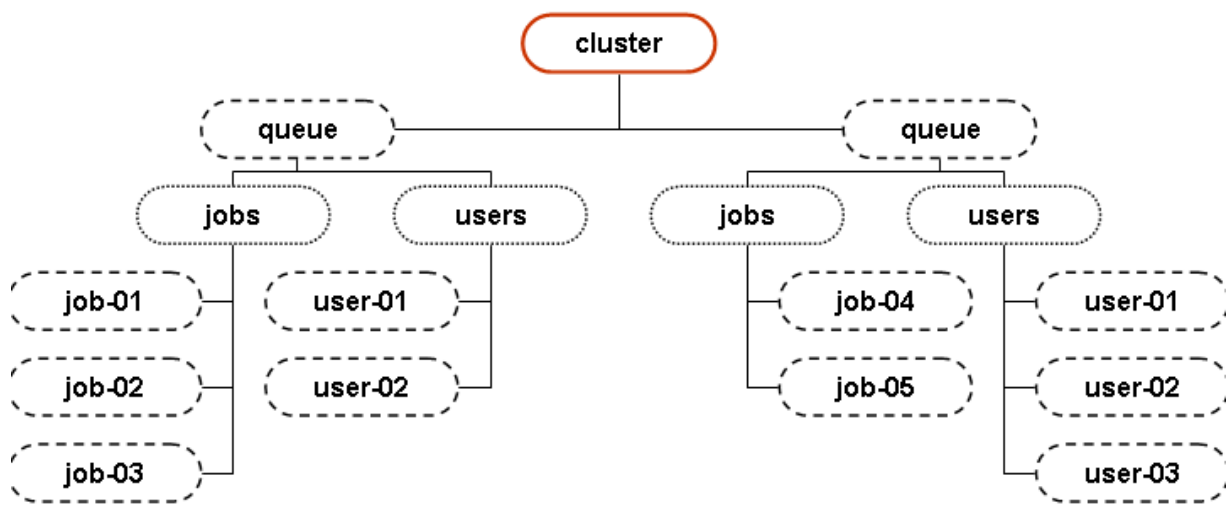}
\caption{\label{fig:dit}The LDAP subtree corresponding to a cluster
resource.} }
\end{figure*}

The NorduGrid model~\cite{is} naturally describes the main Grid
components: computing clusters, Grid users and Grid jobs.  The
LDAP-tree of a cluster is shown in Fig.~\ref{fig:dit}.  In this model,
a \emph{cluster} entry describes the hardware, software and
middleware properties of a cluster.  Grid-enabled queues are
represented by the \emph{queue} entry, under which the
\emph{authuser} and \emph{job} entries can be
found. Every authorized Grid user has an entry under the
queue. Similarly, every Grid job submitted to the queue is represented
by a \emph{job} entry. The job entries are generated on the
execution cluster, this way implementing a distributed job status
monitoring system. The schema also describes Storage Elements and
Replica Catalogs, although in a simplistic manner.

\textbf{The information providers} are small programs that generate
LDAP entries upon a search request. The NorduGrid information model
requires its own set of information providers, therefore a set of such
programs which interface to the local system was created. The
NorduGrid providers create and populate the NorduGrid LDAP entries of
the local database by collecting information from the cluster's batch
system and the Grid Manager, among others, on the status of grid jobs,
grid users and queuing system.

\textbf{The local databases} in the MDS-based system are responsible
for implementing the first-level caching of the providers' output and
answering queries by providing the requested Grid information to the
clients through the LDAP protocol. Globus created a LDAP back-end for
this purpose called the GRIS (Grid Resource Information
Service). NorduGrid uses this back-end as its local information
database. Local databases are configured to cache the output of the
NorduGrid providers for a certain period.


\textbf{Soft-state registration} mechanism of MDS is used by the local
databases to register their contact information into the \textbf{registry
services} (indices) which themselves can further register to other
registries. The soft-state registration makes the Grid dynamic,
allowing the resources to come and go. Utilization of the soft-state
registration made it possible to create a specific topology of indexed
resources.

\textbf{Registries} (or index services) are used to maintain dynamic
lists of available resources, containing the contact information of
the soft-state-registered local databases. It is also capable of
performing queries by following the registrations and caching the
results of the search operations (higher-level caching
mechanism). The Globus developed backend for the registries is called GIIS
(Grid Information Index Service). The higher-level
caching is not used by the NorduGrid Project, and index services are working
as simple dynamic "link catalogs", thus reducing the overall load on
the system. In the NorduGrid system, clients connect the index
services only to find out the contact information of the local
databases (or lower level index services), then the resources are
queried directly.

The local databases and index services of the NorduGrid testbed are organized
into a multi-level tree hierarchy. It attempts to follow a natural
geographical hierarchy, where resources belonging to the same country are
grouped together and register to the country's index service. These
indices are further registering to the top level NorduGrid index
services. In order to avoid any single point of failure, NorduGrid
operates a multi-rooted tree with several top-level indices.

\subsection{User Interface and resource brokering} \label{sec:ui}

The user interacts with the NorduGrid through a set of command line
tools. There are commands for submitting a job, for querying the status
of jobs and clusters, for retrieving the data from finished jobs, for
killing jobs \emph{etc}.  There are also tools that can be used for
copying files to, from and between the Storage Elements and Replica
Catalogs and for removing files from them. The full list of commands is
given below:
\begin{description}
\item{\textbf{\texttt{ngsub}}} -- job submission
\item{\textbf{\texttt{ngstat}}} -- show status of jobs and clusters
\item{\textbf{\texttt{ngcat}}} -- display stdout or stderr of a running job
\item{\textbf{\texttt{ngget}}} -- retrieve the output from a finished job
\item{\textbf{\texttt{ngkill}}} -- kill a running job
\item{\textbf{\texttt{ngclean}}} -- delete the output from the cluster
\item{\textbf{\texttt{ngsync}}} -- recreate the user interface's local information about running jobs
\item{\textbf{\texttt{ngcopy}}} -- copy files to, from and between Storage Elements and Replica Catalogs
\item{\textbf{\texttt{ngremove}}} -- delete files from Storage
  Elements and Replica Catalogs
\end{description}
More detailed information about these commands can be found in the
User Interface user's manual, distributed with the NorduGrid
toolkit~\cite{nordugrid}.

When submitting a Grid job using \texttt{ngsub}, the user should
describe the job using extended RSL (xRSL) syntax
(Section~\ref{sec:xrsl}). This piece of xRSL should contain all the
information needed to run the job (the name of the executable, the
arguments to be used, \emph{etc}.) It should also contain a set of
requirements that a cluster must satisfy in order to be able to run
the job. The cluster can e.g.\ be required to have a certain amount of
free disk space available or have a particular set of software
installed.

When a user submits a job using \texttt{ngsub}, the User Interface
contacts the Information System (see Section~\ref{sec:infosys}): first
to find available resources, and then to query each available cluster
in order to do requirement matching. If the xRSL specification states
that some input files should be downloaded from a Replica Catalog, the
Replica Catalog is contacted as well, in order to obtain information
about these files.

The User Interface then matches the requirements specified in the xRSL
with the information obtained from the clusters in order to select a
suitable queue at a suitable cluster. If a suitable cluster is found,
the job is submitted to that cluster. Thus, the resource brokering
functionality is an integral part of the User Interface, and does not
require an additional service.

\subsection{Resource specification language} \label{sec:xrsl}

The NorduGrid architecture uses Globus RSL~1.0~\cite{gram} as the
basis for the communication language between users, UI and GM. This
extended functionality requires certain extensions to RSL. This
concerns not only the introduction of new attributes, but also the
differentiation between the two levels of the job options
specifications:

\begin{itemize}
\item \textbf{User-side RSL} -- the set of attributes specified by a
user in a job description file. This file is interpreted by the User
Interface (Section~\ref{sec:ui}), and after the necessary modifications
is passed to the Grid Manager (GM, Section~\ref{sec:gm})
\item \textbf{GM-side RSL} -- the set of attributes pre-processed by
the UI, and ready to be interpreted by the GM
\end{itemize}

This dual purpose of the RSL in the NorduGrid architecture, as well as
re-defined and newly introduced attributes, prompted NorduGrid to
refer to the used resource specification language as "xRSL", in order
to avoid possible confusion. xRSL uses the same syntax conventions as
the core Globus RSL, although changes the meaning and interpretation
of some attributes. For a detailed description, refer to the xRSL
specifications distributed with the NorduGrid
toolkit~\cite{nordugrid}.

Most notable changes are those related to the file movement. The major
challenge for NorduGrid applications is pre- and post-staging of
considerable amount of files, often of a large size. To reflect this,
two new attributes were introduced in xRSL: \emph{inputFiles} and
\emph{outputFiles}. Each of them is a list of local-remote file name or
URL pairs. Local to the submission node input files are uploaded to the
execution node by the UI; the rest is handled by the GM. The output
files are moved upon job completion by the GM to a specified
location (Storage Element). If no output location is specified, the
files are expected to be retrieved by a user via the UI.

Several other attributes were added in xRSL, for convenience of users. A
typical xRSL file is:

{\small
\verb#&(executable="ds2000.sh")#\\
\verb#(arguments="1101")#\\
\verb#(join="yes")#\\
\verb#(rsl_substitution=#\\
\verb# ("TASK" "dc1.002000.simul"))#\\
\verb#(rsl_substitution=#\\
\verb# ("LNAM"#\\
\verb#  "dc1.002000.simul.01101.hlt.pythia_jet_17"))#\\
\verb#(stdout=$(LNAM).log)#\\
\verb#(inputfiles=("ds2000.sh"#\\
\verb# http://www.nordugrid.org/$(TASK).NG.sh))#\\
\verb#(outputFiles=#\\
\verb#($(LNAM).log#\\
\verb# rc://dc1.uio.no/2000/log/$(LNAM).log)#\\
\verb#(atlas.01101.zebra#\\
\verb# rc://dc1.uio.no/2000/zebra/$(LNAM).zebra)#\\
\verb#(atlas.01101.his#\\
\verb# rc://dc1.uio.no/2000/his/$(LNAM).his)#\\
\verb#($(LNAM).AMI#\\
\verb# rc://dc1.uio.no/2000/ami/$(LNAM).AMI)#\\
\verb#($(LNAM).MAG#\\
\verb# rc://dc1.uio.no/2000/mag/$(LNAM).MAG))#\\
\verb#(jobname=$(LNAM))#\\
\verb#(runTimeEnvironment="DC1-ATLAS")#\\
}

A more detailed explanation can be found in~\cite{nordugrid-atlas}.

Such an extended RSL appears to be sufficient for job description of desired
complexity. The ease of adding new attributes is particularly appealing, and
NorduGrid is committed to use xRSL in further development.

\subsection{Monitoring} \label{sec:monitor}

The NorduGrid provides an easy-to-use monitoring tool, realized as a Web
interface to the NorduGrid Information System. This Grid Monitor allows
browsing through all the published information about the system,
providing thus a real-time monitoring and a primary debugging for the
NorduGrid.

The structure of the Grid Monitor to great extent follows that of the
NorduGrid Information System. For each class of objects, either an
essential subset of attributes, or the whole list of them, is
presented in an easily accessible inter-linked manner. This is
realized as a set of windows, each being associated with a
corresponding module.

\begin{figure*}
\centering{
\includegraphics[width=0.9\linewidth]{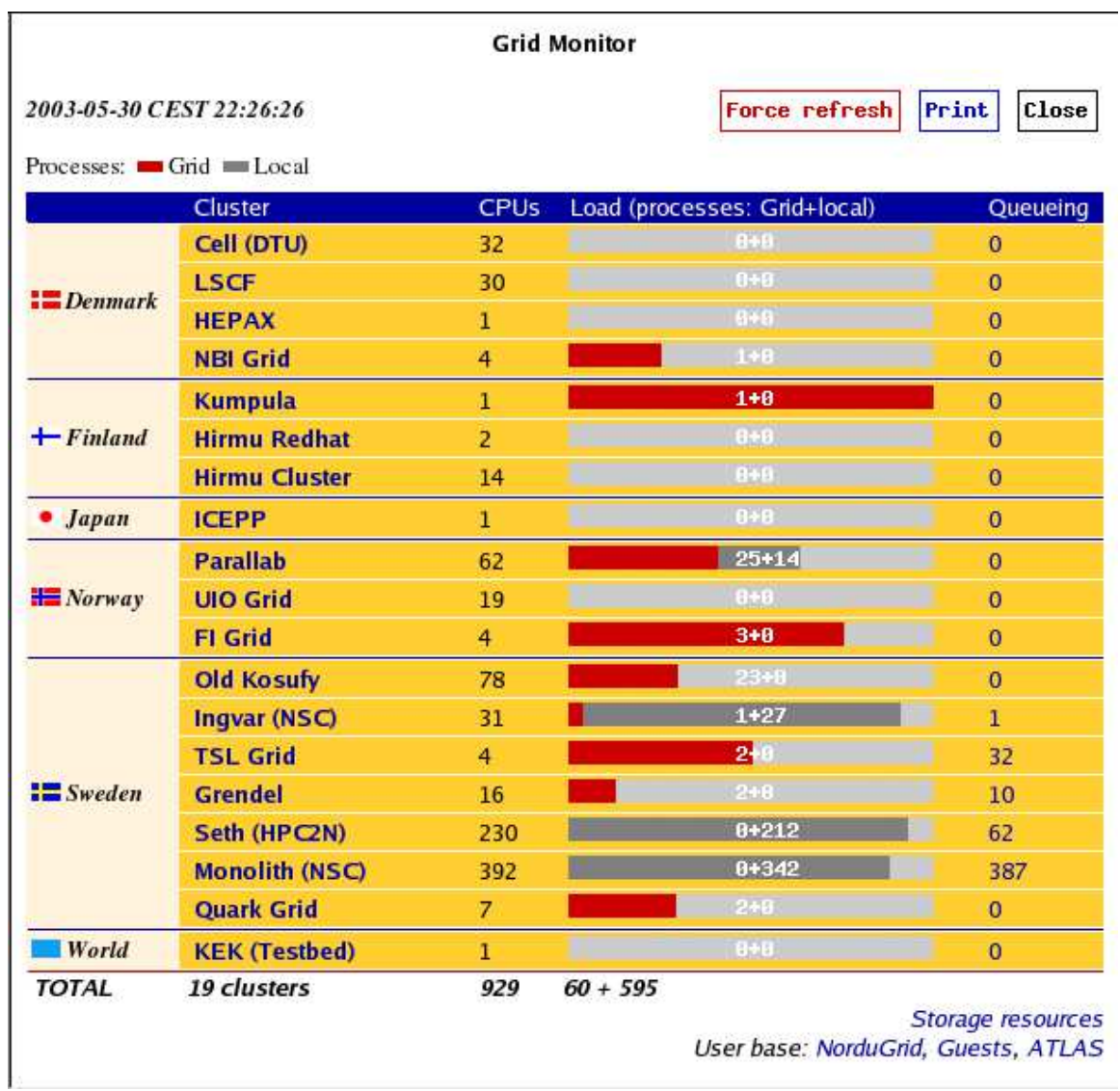}
\caption{\label{fig:loadmon}The Grid Monitor.}}
\end{figure*}

The screen-shot of the main Grid Monitor window, as available from the
NorduGrid Web page, is shown in Fig.~\ref{fig:loadmon}. Most of the
displayed objects are linked to appropriate modules, such that with a
simple mouse click, a user can launch another module window, expanding
the information about the corresponding object or attribute. Each such
window gives access to other modules in turn, providing thus a rather
intuitive browsing.

The Web server that provides the Grid Monitor access runs on a
completely independent machine, therefore imposing no extra load on the
NorduGrid, apart of very frequent LDAP queries (default refresh time for
a single window is 30 seconds).

\subsection{Software configuration and distribution\label{sec:config}}

Since the Grid is supposed to be deployed on many sites, it implies
involvement of many site administrators, of which not all may be Grid
experts. Therefore the configuration of the NorduGrid Toolkit was made
as simple as possible. It basically requires writing two configuration
files: \texttt{globus.conf} and \texttt{nordugrid.conf}. The
\texttt{globus.conf} is used by the \emph{globus-config} package,
which configures the information system of the \globus \ from a single
file. This configuration scheme was developed as a joint effort of
NorduGrid and EDG and thus is not NorduGrid-specific. The
\texttt{nordugrid.conf} file is used for configuring the various
NorduGrid Toolkit components.

This approach proved to be very convenient and allowed to set up sites
as remote from Scandinavia as Canada or Japan in a matter of hours, with
little help from the NorduGrid developers.

The NorduGrid Toolkit is freely available via the Web
site~\cite{nordugrid} as RPM distributions, source tar-balls as well as
CVS snapshots and nightly builds. Furthermore, there is a stand-alone
local client installation, distributed as a tar-ball and designed to be
a NorduGrid entry point, working out-of-the-box. Since it contains the
necessary Globus components, it can be used to interact with other Grids
as well.

\begin{acknowledgments}
The NorduGrid project was funded by the Nordic Council of Ministers
through the Nordunet2 programme and by NOS-N. The authors would like to
express their gratitude to the system administrators across the Nordic
countries for their courage, patience and assistance in enabling the
NorduGrid environment. In particular, our thanks go to Ulf
Mj\"ornmark and Bj\"orn Lundberg of Lund University, Bj\"orn Nilsson of
NBI Copenhagen, Niclas Andersson and Leif Nixon of NSC Link\"oping,
\AA ke Sandgren of HPC2N Ume\aa\ and Jacko Koster of Parallab, Bergen.
\end{acknowledgments}

\end{document}